\documentclass[letterpaper,twocolumn,english,preprintnumbers,amsmath,amssymb,superscriptaddress,nofootinbib,prl]{revtex4}
\usepackage{mathpazo}
\usepackage[T1]{fontenc}
\usepackage[latin9]{inputenc}
\usepackage{verbatim}
\usepackage{amsmath}
\usepackage{amsthm}
\usepackage{amssymb}
\usepackage{graphicx}
\usepackage{setspace}

\makeatletter

\pdfpageheight\paperheight
\pdfpagewidth\paperwidth

\@ifundefined{textcolor}{}
{%
 \definecolor{BLACK}{gray}{0}
 \definecolor{WHITE}{gray}{1}
 \definecolor{RED}{rgb}{1,0,0}
 \definecolor{GREEN}{rgb}{0,1,0}
 \definecolor{BLUE}{rgb}{0,0,1}
 \definecolor{CYAN}{cmyk}{1,0,0,0}
 \definecolor{MAGENTA}{cmyk}{0,1,0,0}
 \definecolor{YELLOW}{cmyk}{0,0,1,0}
}
  \theoremstyle{definition}
  \newtheorem{defn}{\protect\definitionname}
  \theoremstyle{plain}
  \newtheorem{assumption}{\protect\assumptionname}
\theoremstyle{plain}
\newtheorem{thm}{\protect\theoremname}
  \theoremstyle{plain}
  \newtheorem{cor}{\protect\corollaryname}
  \theoremstyle{remark}
  \newtheorem{rem}{\protect\remarkname}
  \theoremstyle{remark}
  \newtheorem*{rem*}{\protect\remarkname}

\usepackage{babel}
\providecommand{\assumptionname}{Assumption}
  \providecommand{\definitionname}{Definition}
  \providecommand{\remarkname}{Remark}
\providecommand{\corollaryname}{Corollary}
\providecommand{\theoremname}{Theorem}

\makeatother

\usepackage{babel}
  \providecommand{\assumptionname}{Assumption}
  \providecommand{\definitionname}{Definition}
  \providecommand{\remarkname}{Remark}
\providecommand{\corollaryname}{Corollary}
\providecommand{\theoremname}{Theorem}

\begin{document}

\title{Generic Local Hamiltonians are Gapless}

\author{Ramis Movassagh}
\email{q.eigenman@gmail.com}
\affiliation{Department of Mathematics, IBM TJ Watson Research Center, Yorktown
Heights, NY, 10598}
\begin{abstract}
We prove that generic quantum local Hamiltonians are gapless. In fact, we prove that there is a continuous
density of states above the ground state. The
Hamiltonian can be on a lattice in any spatial dimension or on a graph
with a bounded maximum vertex degree. The type of interactions allowed
for include translational invariance in a disorder (i.e., probabilistic)
sense with some assumptions on the local distributions. Examples include many-body localization and random spin models. We calculate
the scaling of the gap with the system's size when the
local terms are distributed according to a Gaussian $\beta-$orthogonal
random matrix ensemble. As a corollary there exist finite size partitions
with respect to which the ground state is arbitrarily close to a product
state. When the local eigenvalue distribution
is discrete, in addition to the lack of an energy gap in the limit, we prove that the
ground state has finite size degeneracies. The proofs are simple and constructive. This work excludes the important class of truly translationally
invariant Hamiltonians where the local terms are all equal.
\end{abstract}
\maketitle
The gap of a quantum Hamiltonian is the positive difference of the
two smallest distinct energies. Since finite systems have a finite
number of energy levels, one says that that the system is gapless
if the gap goes to zero in the limit where the size of the system becomes
arbitrarily large. Otherwise, we say that the system is gapped. There is a
stronger notion of gaplessness in which there is a continuous density of
states above the ground state. 

The size of the energy gap has fundamental implications for the physics and simulation 
of quantum many-body systems. In particular, the vanishing of the gap
is a necessary condition for criticality and quantum phase transitions,
while gapped systems exhibit massive excitations and short-range correlations.
 From a quantum computing perspective gapped systems are believed to be easier to classically simulate \cite{landau2015polynomial,verstraete2015worth}. Demonstrating whether a quantum system is gapped or not continues to be a central challenge in condensed matter and quantum information communities. In particular, the spectral gap problem is undecidable \cite{cubitt2015undecidability}.

In mathematics \textit{generic} means almost surely, or with probability
one. Generic, therefore, indicates typical behavior.  Much is known about generic Hermitian operators. For
example, the energy statistics of Gaussian orthogonal ensembles universally
follows the Wigner-Dyson semicircle distribution \cite{hiai2006semicircle}.
The behavior of extreme eigenvalues is given by the Tracy-Widom law
\cite{tracy1994level}. Despite these exciting advances, the standard
results in random matrix theory are of limited applicability in disordered
quantum matter. The key difference is that in quantum many-body systems
the interactions are often \textit{local}. Therefore, generic Hamiltonians reside in low dimensional submanifolds of the Hilbert space. In the physical submanifold far less is known about the generic aspects of
entanglement and energy statistics. 

Generic local Hamiltonians model diverse phenomena in physics from two-dimensional quantum gravity to spin glass phases to the many-body localization problem \cite{binder1986spin,guhr1998random,mehta2004random, pal2010many}. More structured randomness, such as random local projectors
are central in quantum computation and complexity. They are the random clauses in the quantum satisfiability problem, 
and they provide a natural representation for frustration free systems \cite{bravyi2011efficient,movassagh2010unfrustrated}. 

On the one hand, a vanishing gap often requires fine-tuning (say to a critical point).  On the other, Griffith's singularities in disordered
systems may lead to criticality driven by rare regions \cite{cardy1996scaling}.
The occurrence of rare regions is often assumed to be exponentially
small (compare with Remark \ref{rem:scaling}).
These opposing ideas on the generic behavior of the gap date back
at least fifty years. 

So these ideas raise the questions: Can one find a computable low measure
subset of Hamiltonians for which the gap is decidable? Among
all possible quantum local Hamiltonians, is the gapped subset of a full
measure? 

Let  $n$ be the total number of particles and $d$ be the dimension of the local Hilbert space (e.g., number of spin states). Formally, the most general local Hamiltonian acting on the Hilbert
space $(\mathbb{C}^{d})^{\otimes n}$, is
\begin{equation}
H(N)=\sum_{\langle i,j\rangle}\mathbb{I}\otimes H_{i,j}\label{eq:H}
\end{equation}
where $N$ is the total number of summands, $\langle i,j\rangle$
denotes nearest neighbors, $H_{i,j}$'s are $d^{2}\times d^{2}$ Hermitian matrices
and $\mathbb{I}$ is an identity matrix of size $(\mathbb{C}^{d})^{\otimes\left(n-2\right)}$ acting on all sites excluding $i$ and $j$.  

For example, each $H_{ij}$ has $d^{2}+\beta d^{2}(d^{2}-1)/2$
real parameters, where $\beta=1,2,$ and $4$ correspond to real, complex,
and quaternion Gaussian matrices. 
Equation \eqref{eq:H} for a generic spin-$1/2$ chain ($d=2$)
over the complex numbers, in the standard Pauli basis, can be expressed
as (for an open chain $N=n-1$)
\[
H(n)=\sum_{j=1}^{n-1}\mathbb{I}_{2^{j-1}}\otimes\left\{ \sum_{\alpha,\beta=0}^{3}J_{\alpha,\beta,j}\text{ }\sigma_{j}^{(\alpha)}\sigma_{j+1}^{(\beta)}\right\} \otimes\mathbb{I}_{2^{n-j-1}},
\]
where $J_{\alpha,\beta,j}$ are $16n$ random parameters,
and $\sigma^{(0)},\sigma^{(1)},\sigma^{(2)}$ and $\sigma^{(3)}$
are the Pauli matrices. 

Such random Hamiltonians model doped silicon with magnetic impurities, and they are also called random spin exchange models; the low energy physics of these models have been studied in the context of renormalization group theory\cite{fisher1994random,bhatt1982scaling,westerberg1997low}.

In this Letter, we prove under certain assumptions that $H$,
as defined by Eq. \eqref{eq:H}, is with probability one gapless in the thermodynamical
limit. In fact we prove that there is a continuous density of states
above the ground state. The Hamiltonian can be on a lattice
in any spatial dimension or on a graph with bounded vertex degrees. The
results include local Hamiltonians with translational invariance in
a disorder (i.e., probabilistic) sense, where local terms can be independent
random matrices (e.g., spin glass interactions). The origin of gaplessness are local regions with arbitrary small energies. Therefore, the gapless modes are localized excitations unlike what one encounters in the study of metals.

The organization of this paper is as follows. Under Assumption \eqref{assu:jointDensity}, Theorem \eqref{thm:Id_eigenvals} proves a continuous density of states above the ground state. We calculate the scaling of the gap with the system size for local terms that are distributed according to the Gaussian ensemble. Corollary \eqref{cor:discreteEigs} deals with the degeneracy and entanglement of the ground states. We then relax Assumption \eqref{assu:jointDensity}.  Theorem \eqref{thm:Projectors} proves gaplessness for the random local projector problem, and Corollary \eqref{cor:GAPdiscreteEigs-1} for Hamiltonians whose local terms have discrete eigenvalue and Haar eigenvector distributions. In the Supplementary Material \cite{SM} we detail the proofs and derivations.

In the theory of Anderson localization, weak randomness leads to the localization
of eigenstate and the existence of Lifshitz tails \cite{yaida2016instanton}.
In condensed matter physics it is intuitively expected that localization
type interactions would lead to gaplessness \cite{stolz2011introduction}.
Perhaps much more surprising is the gaplessness of general disordered
systems given by Eq. \eqref{eq:H} that among other physical systems
include quantum spin systems as well. 
\begin{rem}\label{rem:degVsGap}We make a sharp distinction between finite
size degeneracies and gaplessness. Therefore, here
the gap is always positive for finite size systems.
\end{rem}
In Eq. \eqref{eq:H} to meaningfully take a limit of an arbitrary large
$N$, we assume that $H_{i,j}$'s have bounded operator norms, i.e., a bounded maximum singular value. We also assume that the number of neighbors
of any given $H_{i,j}$ is a constant independent of $N$. We work
with finite dimensional Hilbert spaces and consider $N$ to be arbitrarily
large. Below, for simplicity, we denote $H(N)$ simply by $H$ and say
matrices $A$ and $B$ are $\epsilon$ close if $\left\Vert A-B\right\Vert \le\epsilon$.

The Hamiltonian induces a probability measure, or a density of states
$\mu[H(N)]$, over the real numbers, such that the integral over an
interval gives the expected fraction of the eigenvalues in it. Formally,
\[
d\mu(\lambda)=\frac{1}{d^{n}}\left\langle \sum_{i=1}^{d^{n}}\delta\left(\lambda-\lambda_{i}\right)\right\rangle \text{ }d\lambda.
\]
The following gapless definition only requires that the
two smallest distinct eigenvalues merge as $N\rightarrow\infty$.
\begin{defn}
\label{Def:Weak} (Weak) Let $\lambda_{0}(H)$ and $\lambda_{1}(H)$
be the smallest and second smallest distinct eigenvalues of $H$ respectively.
The gap of $H$ is defined to be $\lambda_{1}(H)-\lambda_{0}(H)$.
We say $H$ is gapless if for any constant $\epsilon>0$ there exists $N>0$
such that $\lambda_{1}(H)-\lambda_{0}(H)\le\epsilon$. 
\end{defn}
There is yet a stronger definition in which there is a continuous
density of states connected to and above the ground state; please
see Fig. \eqref{fig:Notions_of_Gap} and \cite{SM} for a precise
definition. 

{\label{sec:Continuous-local-terms}{\it Continuous local terms }}--In this section, we assume that the joint distribution of the eigenvalues
of a local term obeys a niceness property that most standard finite
random matrix ensembles, such as Gaussian orthogonal ensembles, possess
\cite{mehta1983spacing,marvcenko1967distribution}. Mathematically, 
\begin{assumption}
\label{assu:jointDensity} For all $\epsilon>0$, there exists a real
number $\mu_{i,j}$ such that the probability for all of the eigenvalues
of $H_{i,j}$ to be within $\epsilon$ of $\mu_{i,j}$ is positive. 
\end{assumption}
Let us illustrate Assumption \ref{assu:jointDensity} by constructing
a local term from the Gaussian unitary ensemble (GUE). Let $A$ be
a $d^{2}\times d^{2}$ matrix whose entries are standard complex normals.
Then $(A+A^{\dagger})/2$ is an instance of GUE whose eigenvalues
we claim can be arbitrarily close. In particular, $A$ has a positive
probability of being $\epsilon$ close to a multiple of the identity.

\begin{figure}
\begin{centering}
\includegraphics[scale=0.3]{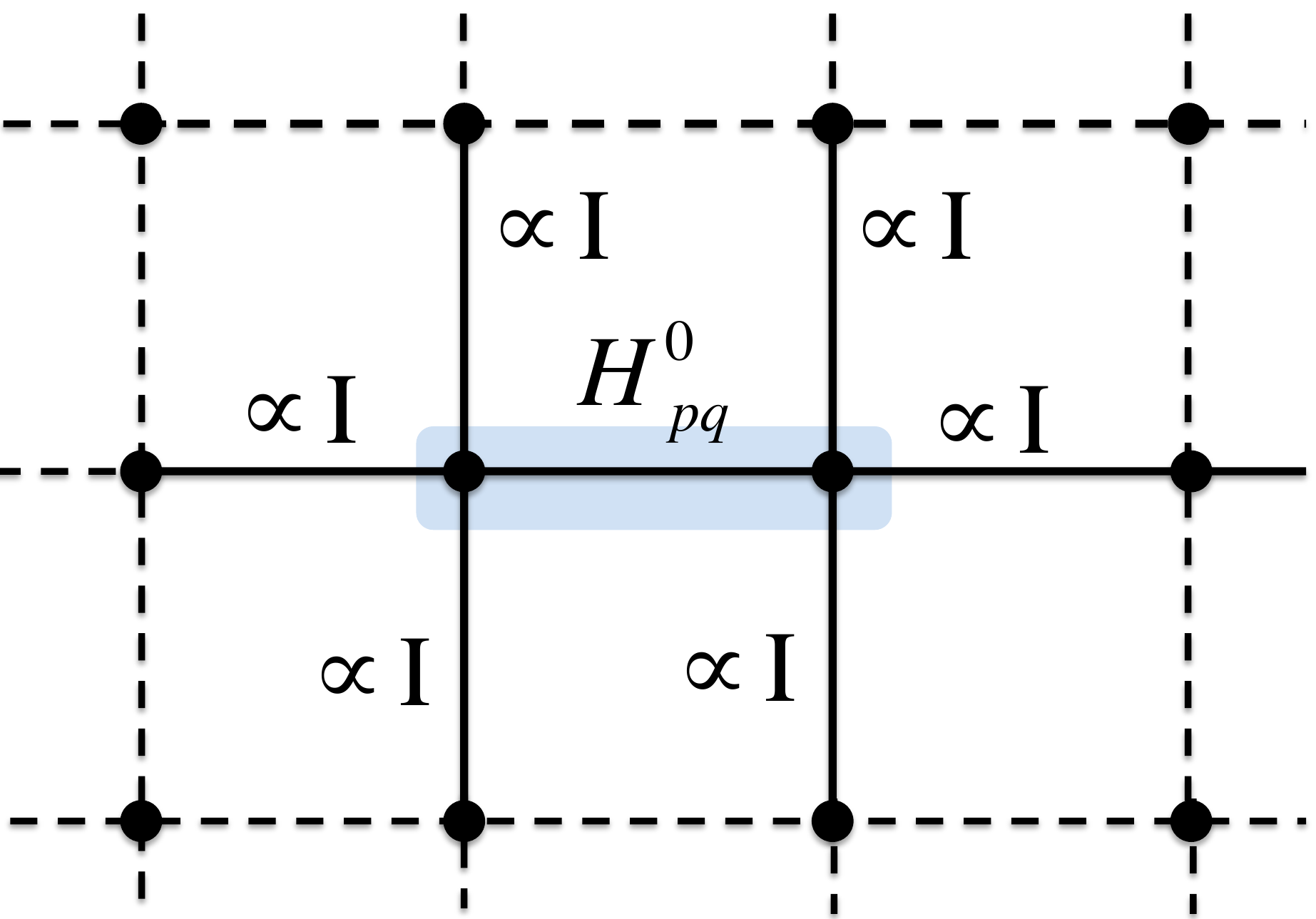}\caption{\label{fig: Theorem1}
Illustration of a rare local region (e.g.,  $H_{0}$ in Eq. \eqref{eq:rareLocal1} in \cite{SM}), where all distance $1$ terms
are shown in solid lines, and terms with a distance greater than $1$
are shown in dashed lines. }
\par\end{centering}
\end{figure}

Since they are simple and constructive, below we give intuitive explanations of the proofs and defer the rigorous proofs to  \cite{SM}
\begin{thm}
\label{thm:Id_eigenvals}$H$ as defined in Eq.\eqref{eq:H} is almost surely gapless and has a continuous
density of states above the ground state, if $H_{i,j}$'s are independent
and each $H_{i,j}$ has a continuous joint distribution of eigenvalues
that obeys Assumption \ref{assu:jointDensity}.
\end{thm}
Because of Assumption \ref{assu:jointDensity} there is a positive
probability that there exists a $H_{p,q}$ whose two smallest eigenvalues
are $\epsilon$ apart (see Fig.\ref{fig: Theorem1}). In addition, by the independence of the local terms and Assumption \ref{assu:jointDensity},
there is a positive probability that every neighbor of $H_{p,q}$
is $\epsilon-$close to a multiple of the identity. Hence, $H_{p,q}$ essentially becomes a decoupled local region with a gap of size $\epsilon$. And $\epsilon$ can be arbitrary small. 

For the second part, suppose such a localized region is realized for a given system size, we then proceed by making the system size larger and larger to have more and more such local regions. This would ensure a monotonically increasing number of eigenvalues that all are $\epsilon-$close to the ground state (see \cite{SM} for the proof).

The scaling of the gap with the systems' size gives insight
into the physical properties such as the dynamical exponents. It generally would depend
on the local covariance matrix, the parameters and symmetries of the Hamiltonian. Therefore, it would be impossible
to calculate the gap scaling without further information.

The $\beta-$Gaussian ensemble is the canonical random matrix distribution, where
$\beta=1,2,$ and $4$ refer to orthogonal (GOE), unitary
(GUE) and symplectic (GSE) ensembles respectively. We now calculate
the scaling of the gap with the system's size for local terms that are independently drawn from the $\beta-$Gaussian ensemble. 
 \vspace{-5mm}
\subsection{\label{subsec:GUE_GOE_GSE}Example: Scaling of the gap when local
terms are from canonical Gaussian ensembles}
We first answer the following question: What is the probability that
all the eigenvalues of an instance of an $n\times n$ G(O/U/S)E matrix
are $\epsilon-$close? In other words, what is the probability that
a random matrix, $M_n$, from the Gaussian $\beta-$ensemble is $\epsilon-$close
to a multiple of an identity? Throughout this work we think of $n$ as a fixed
positive integer. A lot is known about asymptotic behavior, but for
the problem at hand, $n=d^{2}$ can be quite small. 

Mathematically, for $0<\epsilon\ll1$ we want to calculate the following
probability 
\begin{equation}
\text{P}\left[\exists\mbox{ } a\in\mathbb{R}: \left\Vert M_{n}-a\text{\ensuremath{\mathbb{I}}}_{n}\right\Vert _{F}\le\epsilon\right],\label{eq:ProblemStatement}
\end{equation}
where $\left\Vert \centerdot\right\Vert _{F}$ is the Frobenius norm,
$\mathbb{I}_{n}$ is an $n\times n$ identity matrix, and $M_{n}$ is
a $n\times n$ matrix from G(O/U/S)E ensemble. In words, this is the
probability that $M_{n}$ is within a ball of radius $\epsilon$ centered
around any multiple of the identity. 

The measure over G(O/U/S)E matrices  is 
\[
\mu_{n}(\beta)=C_{n}(\beta)\text{ }e^{-\frac{\beta}{4}\text{tr}(M_{n})^{2}}dM_{n}.
\]

Let us fix a real number $a$, and compute the probability $\text{P}\left[\left\Vert M_{n}-a\mathbb{I}_{n}\right\Vert _{F}\le\epsilon\right]$. The probability density of $M_{n}$ is proportional to
\[
e^{-\frac{\beta}{4}\text{Tr}(a\mathbb{I}_{n})}=e^{-\frac{\beta na^{2}}{4}}
\]
and the volume of a ball of radius $\epsilon$ in $n^{2}$ dimensions
is proportional to $\epsilon^{n^{2}}$, so (see \cite{SM} for derivation)
\begin{equation}
\text{P}\left[\left\Vert M_{n}-a\mathbb{I}_{n}\right\Vert _{F}\le\epsilon\right]=\left(C'_{n}(\beta)+o(1)\right)\epsilon^{n^{2}}e^{-\frac{\beta na^{2}}{4}},\label{eq:Prob_near_a}
\end{equation}
where $C'_{n}(\beta)$ is a constant that depends on $n$ and $\beta$,
and an $o(1)$ term that goes to zero as $\epsilon$ goes to zero. But $a$
is an arbitrary point on the real line, the contribution from all
such points is upper bound by the (Gaussian) integral over $a$, which also scales as $\epsilon^{n^{2}}$. Hence, the probability that $M_n$ is $\epsilon-$close to a multiple of the identity is $\Theta(\epsilon^{n^2})$.

Suppose the number of overlapping terms with $H_{p,q}$ is $z$. Because
of their independence and the argument above, the probability of all of them being $\epsilon-$close
to a constant multiple of the identity is $\left[K_{n}(\beta)+o(1)\right]\epsilon^{zd^{4}}$,
where $K_{n}(\beta)$ is a constant and we restored  $n=d^{2}$. Moreover the probability that
the smallest two eigenvalues of a matrix chosen from the Gaussian
ensemble are $\epsilon$-close is bounded by $\epsilon^{4}$ \cite[This is true for general Wigner matrices]{ erdHos2009local}.
Therefore the probability that a rare local region
occurs is (see \cite{SM} for details)
\[
\left(K_{d}'(\beta)+o(1)\right)\epsilon^{zd^{4}+4}.
\]
Note that the $\beta$ dependence only enters through the prefactor
that is independent of $\epsilon$. 

How large does the system have to be for the expected gap to be $\epsilon?$
For the gap to be $\epsilon$ small, the expected number of term in
the Hamiltonian Eq. \eqref{eq:H} is $N\sim\epsilon^{-zd^{4}-4}$, and
we conclude that 
\begin{equation}
\epsilon(N)\sim N^{-1/(zd^{4}+4)}.\label{eq:Eps_vs_N}
\end{equation}
The scaling of the size with the gap that we just derived depends on the
particular structure of the rare local region, where the neighboring
sites of $H_{p,q}$ are $\epsilon-$close to a multiple of an identity.
There may be different rare, or perhaps nonlocal configurations that protect the near two-fold
degeneracy of the ground state of $H_{p.q}$ with less restriction
on the overlapping terms. It would be interesting if other configurations
were found that gave the scaling $\epsilon(N)\sim N^{-1/4}$. We leave
this for future works.

We now turn to Eq. \eqref{eq:H} with local terms that have discrete eigenvalue
distributions such as random local projectors of a fixed rank. These violate Assumption \ref{assu:jointDensity}, because the overlapping terms can have zero probability of being close to a multiple of the identity. Moreover, the gap of a single
local projector is always one. These are treated independently in Corollary \ref{cor:GAPdiscreteEigs-1} and Theorem \ref{thm:Projectors}.

{\label{sec:Discrete-local-terms}{\it Discrete local terms}}-- More restrictive are local terms with \textit{discrete }eigenvalue
distributions, in which any two local eigenvalues are either equal
or a constant apart; i.e., cannot be $\epsilon$ close. Also the neighbors
are either exactly a multiple of the identity or a constant distant
apart. Corollary \ref{cor:GAPdiscreteEigs-1} proves that such Hamiltonians
are also gapless. Now we prove that their ground states are degenerate. 
\begin{cor}
\label{cor:discreteEigs}If the local eigenvalues have a discrete
distribution satisfying Assumption \ref{assu:jointDensity}, then
the ground state is almost surely degenerate and can be represented
as a product state. 
\end{cor}
If the discrete distribution has a finite number of atoms in its distribution,
then the above corollary suggests a \textit{large} degeneracy of the
ground state for large but finite $N$. We now turn to the gap of random projectors that does
not obey Assumption \ref{assu:jointDensity}. 
\begin{thm}
\label{thm:Projectors}$H$ is almost surely gapless if the local
terms have independent and Haar distributed eigenvectors, and $H_{i,j}$
in Eq. \eqref{eq:H} are random projectors whose individual ranks
are either: \\
 1. Fixed and at most $d(d-h)$, where $1\le h\le d-1$. \\
 2. Vary randomly among the terms in the Hamiltonian. 
\begin{figure}
\begin{centering}
\includegraphics[scale=0.35]{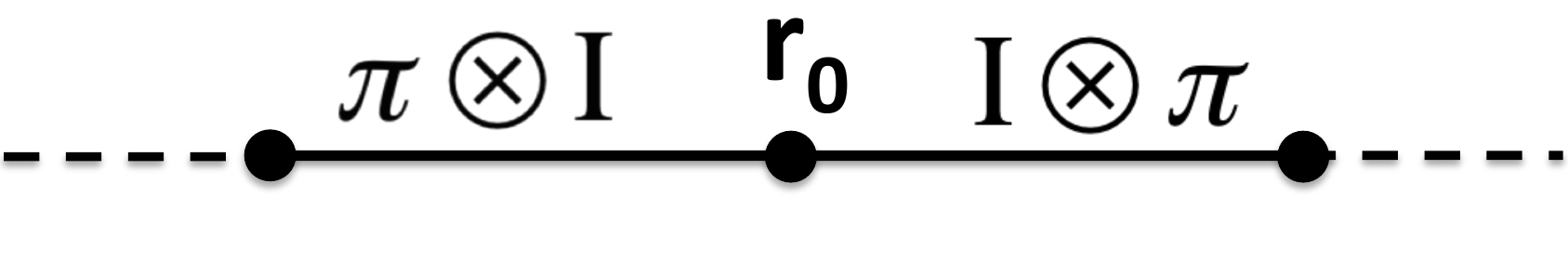}
\par\end{centering}
\begin{centering}
\includegraphics[scale=0.30]{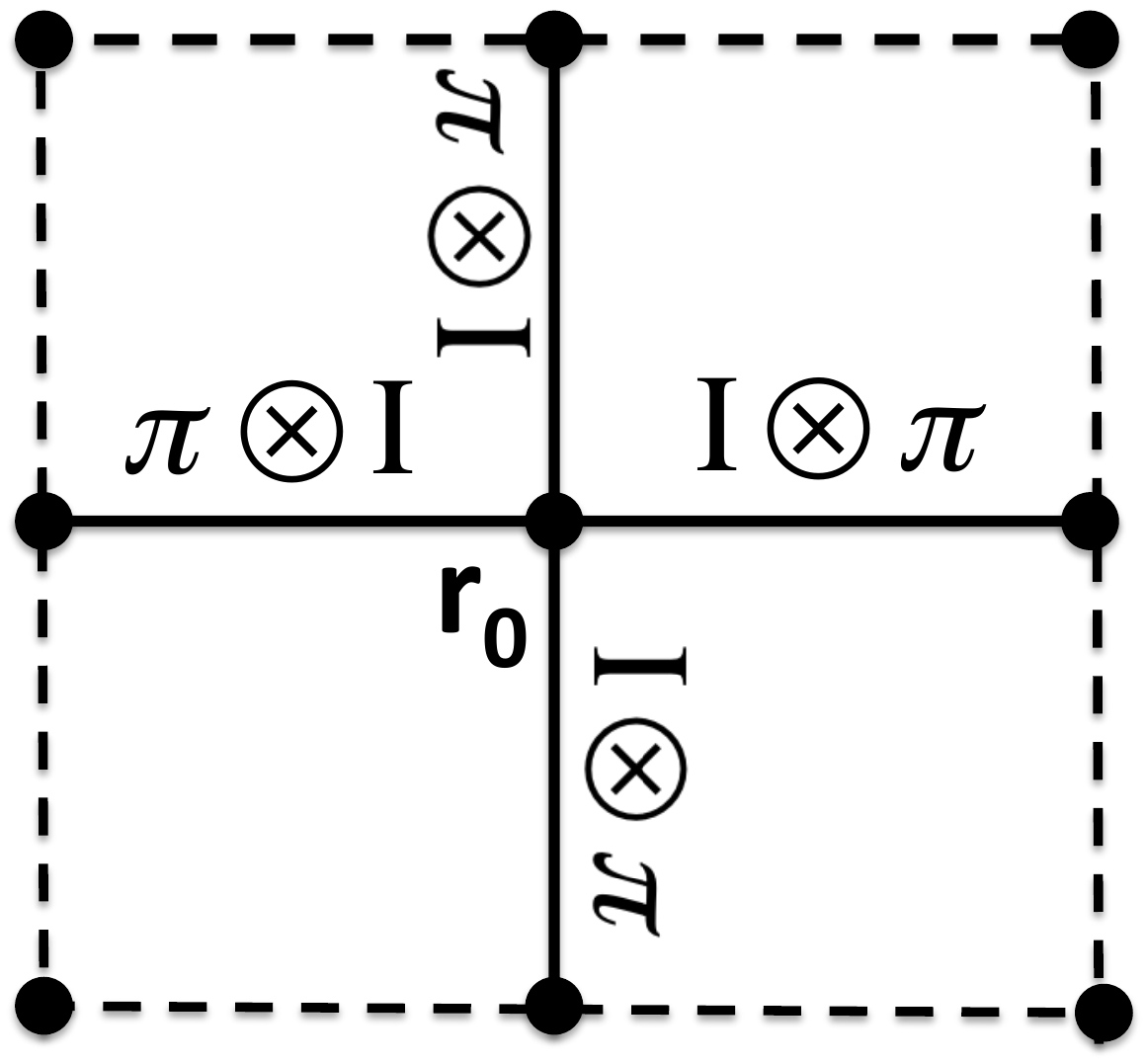}
\par\end{centering}
\centering{}\caption{\label{fig:An-example:-Configuration}Examples of a rare local projector (e.g., $H_{0}$ in Eq. \eqref{eq:H0} in \cite{SM})
on a line and a square lattice, where for simplicity we dropped the
subscript $\mathbf{r_{0}}$ on $\mathbb{I}$}
\end{figure}
\end{thm}
Let $\pi=\mbox{diag}(0,1,\dots,1)$ be the diagonal projector of rank
$d-1$ and $\mathbb{I}_{\mathbf{r_{0}}}$ an $d\times d$ identity
matrix acting on $\mathbf{r_{0}}$. Since eigenvectors are Haar, there
is a positive probability density that a configuration $\epsilon-$close
to those shown in Fig. \eqref{fig:An-example:-Configuration} occurs,
where each local projector $\pi\otimes\mathbb{I}$ has rank $d(d-1)$. Such configurations
basically disconnect a site at $\mathbf{r_{0}}$, which results in
$d-$fold degeneracies of all the eigenvalues of the sum of all the
 interaction that exclude $\mathbf{r_{0}}$. Therefore, when the configuration
is $\epsilon-$close to disconnecting the site at $\mathbf{r_{0}}$,
what used to be exactly a $d-$fold degenerate ground state splits
into $d$ states $\epsilon$ apart for any arbitrarily small $\epsilon$. 

This formally generalizes to local projectors with
ranks $d(d-h)$, with $d>h\ge1$ on lattices or graphs whose degrees
are constants independent of $N$. All that needs to be changed in
the proof is to let $2\le\mbox{dim}[\mbox{Ker}(\pi)]<d$. 

Lastly in the second part of the theorem, where we allow variable ranks among $H_{i,j}$'s, the same construction
as above guarantees having a site on which the Hamiltonian acts trivially
and the above argument guarantees $\epsilon$ splitting of the eigenvalues
and lack of an energy gap in the limit. The proof for having a continuous
density of states above the ground state is in the same spirit as 
the one given for Theorem \ref{thm:Id_eigenvals}. See \cite{SM} for the
formal proof. 
\begin{rem}\label{rem:scaling}
The probability of being $\epsilon$-close to a local
disconnected region depends on the local distribution
of the eigenvectors. It need not be exponentially small as often assumed in Griffith's theory. 
\end{rem}
In Corollary \ref{cor:discreteEigs} we proved that the ground state is degenerate. The following proves that the gap also closes
without using Assumption \ref{assu:jointDensity}; i.e., irrespective of and in addition to possible ground state degeneracies. 
\begin{cor}
\label{cor:GAPdiscreteEigs-1}The Hamiltonian is almost surely gapless
if the local eigenvalues have a discrete eigenvalue distribution and
eigenvectors are Haar distributed.
\end{cor}
This generalizes Theorem \eqref{thm:Projectors} by allowing the local eigenvalues to have more atoms than just zero and one (see \cite{SM} for the proof).

{\label{sec:Discussions-and-conclusions}{\it Limitations, discussions
and conclusions}}--In nature, strict translational invariance is not realistic. However,
to make the study of matter more tractable, idealized models are often
considered that have strict translational symmetry. Our work excludes this
important class of Hamiltonians, where all the local terms are exactly
equal. There are various such models that are gapped (e.g., the AKLT)
or gapless (e.g., critical systems); therefore, there may not be a
unique generic gap behavior.  It might be possible
to classify the gap when the local terms are all equal. This has
been shown recently only for spin-$1/2$ frustration free spin chains \cite{bravyi2015gapped}.

In the proofs we used Weyl's inequalities and \textit{not} perturbation theory. Weyl's inequalities bound the maximum deviation of an eigenvalue of a Hermitian matrix resulting from additive perturbations.\textit{The general issue with using standard perturbation
theory to prove the lack of an energy gap is that one would have to prove
that a constant gap does not open up in all higher order terms}. This
is not obviously so, since the $k^{\mbox{th}}$ correction multiplying
$\epsilon^{k}$ involves a sum of combinatorially many terms and may
actually become comparable to $\epsilon^{-k}$ in the magnitude. This
in turn causes the series to diverge beyond that order. This issue
is often met with in (particle) physics. 

The spectral gap problem is undecidable,
nevertheless we proved, under minimal genericity assumptions, that local Hamiltonians
are gapless in any dimension. 

On the one hand, the extremal eigenvalue statistics of (dense) random matrices
are governed by the Tracy-Widom laws \cite{tracy1994level}. On the
other hand, for classical spin glasses, central limit theorem ensures
Gaussian statistics of the eigenvalues. The density of states of generic
quantum spin chains was shown to be in between the two extremes \cite{movassagh2011density};
it would be interesting to investigate the generic gap scaling
as a function of the two extremes.

Lastly, the existence of rare regions has other implications for the physics of disordered systems. For example, in one dimension, it could lead to the lack of transport by decoupling of the chain into two pieces.
\begin{acknowledgments}
I thank Toby Cubitt, Peter Shor and Barry Simon for discussions. I
thank the AMS-Simons travel grant and IBM T. J. Watson Research Center
for the freedom and support.
\end{acknowledgments}
\bibliography{mybib}
\clearpage
\newpage
\section{\label{sec:Appendix}Supplementary Material: Generic local Hamiltonians are gapless}
\subsection{\label{sec:Appendix:-Weyl-Inequalities}Weyl Inequalities}
Let $H$ be a fixed $n\times n$ Hermitian matrix. Let $\lambda_{0}(H)\le\lambda_{1}(H)\le\dots\le\lambda_{n-1}(H)$ be the eigenvalues of $H$. We denote by $\lambda^\uparrow(H)$ the vector with components $\lambda^\uparrow_j=\lambda_j(H)$, where $0 \le j \le n-1$. Similarly $\lambda^\downarrow$ denotes the vector with components $\lambda_j^\downarrow=\lambda_{n-j-1}$.
\begin{thm}(Weyl) Let $H$ and $V$ be $n\times n$ Hermitian matrices. Then we have\\
$\lambda^\downarrow_j(H+V)\le \lambda^\downarrow_k(H)+\lambda^\downarrow_{j-k}(V)$,  for $k\le j$,\\
$\lambda^\downarrow_j(H+V)\ge \lambda^\downarrow_k(H)+\lambda^\downarrow_{j-k+n-1}(V)$,  for $k\ge j$.
\end{thm}
See Chapter 3 in Bhatia's book for further details \cite{bhatia2013matrix}. 

From this theorem, it is an exercise to prove the following:
\[
\lambda^\downarrow_{j}\left(H\right) - \left\Vert V\right\Vert \le\lambda^\downarrow_{j}\left(H+V\right)\le\lambda^\downarrow_{j}\left(H\right)+\left\Vert V\right\Vert.
\]

In the article and proofs below we mostly applied this to the smallest eigenvalue
(ground state energy). Namely, 
\[
\lambda_{0}\left(H\right) - \left\Vert \delta H\right\Vert \le\lambda_{0}\left(H+\delta H\right)\le\lambda_{0}\left(H\right)+\left\Vert \delta H\right\Vert.
\]

In the proofs we used Weyl's theorem. Had we used first order perturbation
theory, for example in Eq. \eqref{eq:projectors}, instead of $||\sum_{|\langle i,\mathbf{r_{0}}\rangle|=1}\mathbb{I}\otimes\delta H_{i,\mathbf{r_{0}}}||$,
we would have $\lambda_{min}^{\epsilon,k}=\lambda_{E}^{(0)}+\sum_{|\langle i,j\rangle|=1}\langle\theta_{E}\otimes e_{k}|\delta H_{i,\mathbf{r_{0}}}|\theta_{E}\otimes e_{k}\rangle+O(\epsilon^{2})$.
This would not guarantee that a constant small gap would not open in a higher order term in the perturbation expansion.\subsection{Supplementary: Rigorous Results and Proofs}
\begin{figure}
\begin{centering}
\includegraphics[scale=0.35]{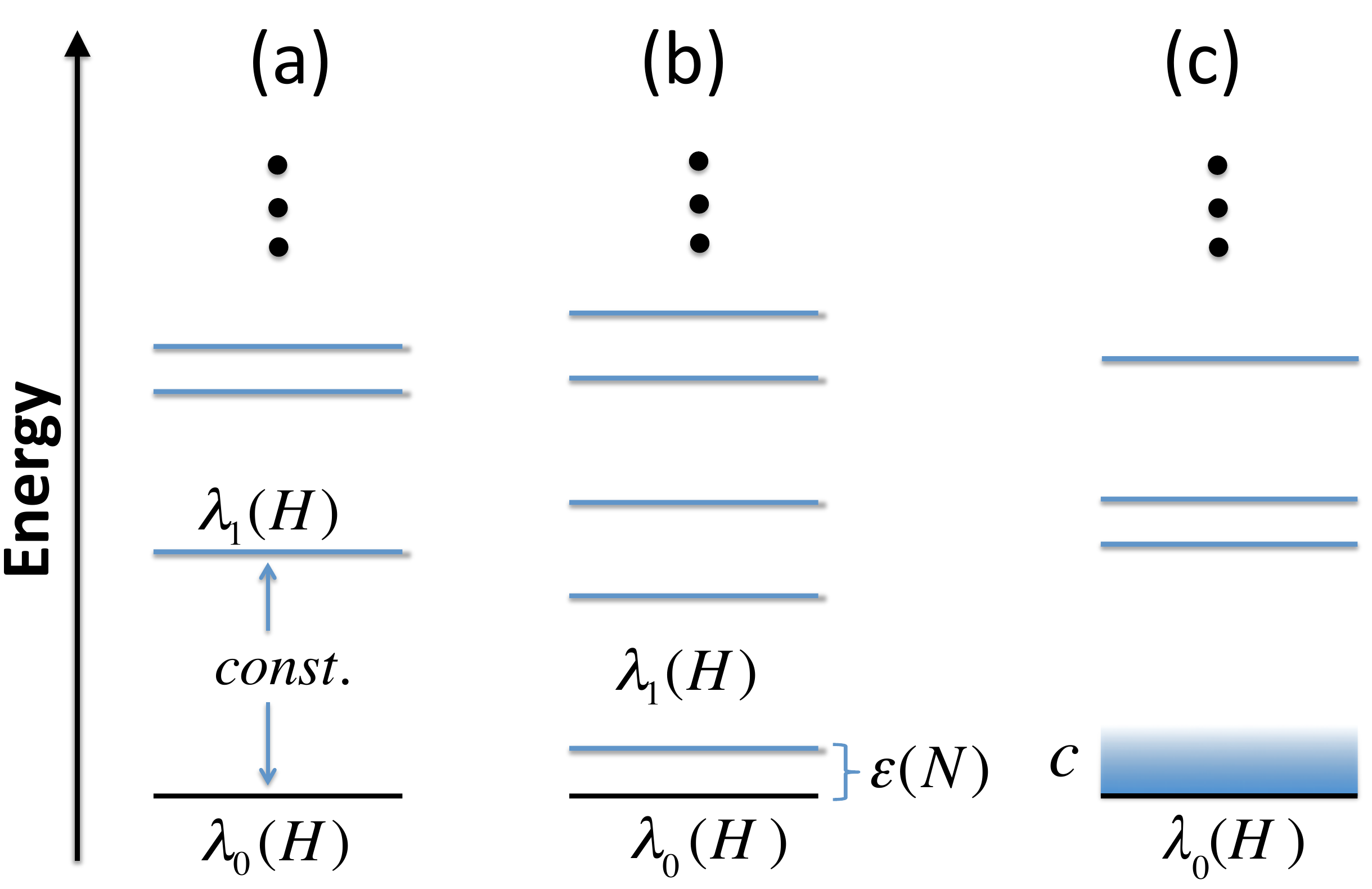}
\par\end{centering}
\caption{\label{fig:Notions_of_Gap}This figure schematically illustrates the
energy levels of the system and what is meant
by the system being a) Gapped, b) Gapless as defined in Definition \eqref{Def:Weak}
(weaker notion) c) Strongly gapless as defined in Definition \eqref{Def:Stronger}; i.e., a continuous density of
states above the ground state.}
\end{figure}

The stronger definition of gaplessness is that there is a continuous
density of states above the ground state. Mathematically, 
\begin{defn}
\label{Def:Stronger} (Continuous Density of States) For any small constant $c>0$, any point $s\in [0,c]$, and $\epsilon\ll c$ , there exist positive integers $j>0$, and $N_j > 0$ such that $\left|\lambda_j-(\lambda_0+s)\right|\le\epsilon$ for all $N>N_j$. See Fig. \ref{fig:Notions_of_Gap}.
\end{defn}
\subsection{Proof of Theorem \ref{thm:Id_eigenvals}}
\begin{proof}
We first prove the weaker version. Let $H_{p,q}$ be a fixed local term and rewrite Eq. \eqref{eq:H} as 
\begin{equation}
H=\mathbb{I}\otimes H_{p,q}+\sum_{|\langle i,j\rangle|=1}\mathbb{I}\otimes H_{i,j}+\sum_{|\langle i,j\rangle|\ge2}\mathbb{I}\otimes H_{i,j}\label{eq:HammingDistance-1}
\end{equation}
where the first sum includes all the local terms that overlap with
$H_{p,q}$ at a site (i.e., have distant $1$) and the second sum
all the terms with no overlap with $H_{p,q}$ (at least a distant
$2$). Let the number of overlapping terms be $z$. For example, on
a square lattice $z=4D-2$ where $D$ is the spatial dimension.

Because of Assumption \ref{assu:jointDensity} there is a positive
probability that there exists $H_{p,q}$ whose two smallest eigenvalues
are $\epsilon$ apart. In addition, by the independence of the local terms
there is a positive probability that every neighbor of $H_{p,q}$
is $\epsilon$ close to a multiple of the identity. At $\epsilon=0$,
we define $H_{0}$ to be (see Fig. \eqref{fig: Theorem1})
\begin{equation}
H_{0}\equiv\mathbb{I}\otimes H_{p,q}^{0}+\sum_{|\langle i,j\rangle|=1}\beta_{i,j}\mathbb{I}\otimes\mathbb{I}_{i,j}+\sum_{|\langle i,j\rangle|\ge2}\mathbb{I}\otimes H_{i,j}\label{eq:rareLocal1}
\end{equation}
where the superscript zero on the first term means that $H_{pq}^{0}$
has a degenerate smallest eigenvalue denoted by $\lambda_{0}$.

Let $\lambda_{E}$ be the smallest eigenvalue of $\sum_{|\langle i,j\rangle|\ge2}\mathbb{I}\otimes H_{i,j}$.
By assumption $H_{p,q}=H_{p,q}^{0}+\delta H_{p,q}$, where $||\delta H_{p,q}||\le\epsilon$ is a small operator (small deviation from $H^0_{p,q}$), 
and the summands of distant 1 terms are $H_{i,j}=\beta_{i,j}\mathbb{I}_{i,j}+\delta H_{ij}$,
where $||\delta H_{i,j}||\le\epsilon$. Since $||\mathbb{I}\otimes\delta H_{i,j}||=||\delta H_{i,j}||$,
by Weyl's inequalities the two smallest eigenvalues of $H$, denoted by $\lambda_{min}^{\epsilon,k}$
with $k\in\{1,2\}$, obey 
\begin{equation}
\lambda_{E}+\beta+\lambda_{0}-B\le\lambda_{min}^{\epsilon,k}\le\lambda_{E}+\beta+\lambda_{0}+B,\label{eq:continuous}
\end{equation}
where $B=||\delta H_{p,q}+\sum_{|\langle i,j\rangle|=1}\delta H_{i,j}||\le\epsilon(z+1)$,
and $\beta\equiv\sum_{|\langle i,j\rangle|=1}\beta_{i,j}$ . Since
for any fixed $\epsilon$ the configuration just described has a positive
probability, we can find a sufficiently large $N_{0}$ such that for
the system's size $N>N_{0}$ there is a site whose two smallest eigenvalues
are $\epsilon$ apart and whose neighbors are $\epsilon$ close to
a multiple of the identity. Since $z+1$ is finite and $\epsilon$
can be arbitrary small, we conclude that $H$ is gapless in the weaker
sense of Definition \eqref{Def:Weak}. This completes the proof of
lack of an energy gap between the ground state and the first excited
state. 

Now suppose the configuration given by Eq. \eqref{eq:rareLocal1} is
realized for $N=N_{0}$, where instead now $\lambda_1(H^0_{p,q})-\lambda_0(H^0_{p,q})=s$, where $s\in[0,c]$ as in Definition \eqref{Def:Stronger}, and $c$ is a constant smaller than or equal to the width of the support of the local eigenvalues. The existence of the interval $[0,c]$ is guaranteed by Assumption \ref{assu:jointDensity}. There is a positive probability that $H_{p,q}$ and each of the overlapping terms are $\epsilon\ll c$ close to the first two terms in Eq. \eqref{eq:rareLocal1} respectively. The application of Weyl inequalities (similar to the weaker proof) guarantees that the point $s$ is $\epsilon-$close to an eigenvalue of $H$.  Since the point $s$ is arbitrary and has a positive probability of being the gap of $H^0_{p,q}$, we are guaranteed that as we make the system size larger and larger, in the limit every point inside the interval $[0,c]$ is $\epsilon-$close to an eigenvalue of $H$. This proves that the density of states is continuous in an interval of size $c$ above the ground state.
\end{proof}
$k-$local Hamiltonians with $k>2$ have a similar proof. 
\subsection{Derivation of the scaling of the gap for GOE, GUE and GSE}
We first answer the following question: What is the probability that
all the eigenvalues of an instance of an $n\times n$ G(O/U/S)E matrix
are $\epsilon-$close? In other words, what is the probability that
a random matrix from the Gaussian $\beta-$ensemble is $\epsilon-$close
to a multiple of identity? Throughout we think of $n$ as a fixed
positive integer. A lot is known about asymptotic behavior, but for
the problem at hand $n$ can be quite small. 

Let $M_{n}$ be a G(O/U/S)E matrix with eigenvalues $\lambda_{n-1}\ge\lambda_{n-2}\ge\dots\ge\lambda_{0}$
; it is well-known that the eigenvalue density is \cite[Section 2.6]{tao2012topics}
\begin{widetext}
\begin{equation}
\rho_{\beta}(\lambda_{0},\lambda_{1},\dots,\lambda_{n-1})=Z_{n}(\beta)\text{ }\exp\left(-\frac{\beta}{4}\sum_{j=0}^{n-1}\lambda_{j}^{2}\right)\quad\Pi_{0\le j<k\le n-1}|\lambda_{j}-\lambda_{k}|^{\beta}\label{eq:GUE_tao-1}
\end{equation}
\end{widetext}
where $Z_{n}(\beta)$ is the normalization constant that only depends
on $\beta$ and $n$. In the GUE case, $Z(\beta=2)=\frac{1}{(2\pi)^{n/2}\Pi_{j=1}^{n-1}j!}$. 

Mathematically, for $0<\epsilon\ll1$ we want to calculate the following
probability 
\begin{equation}
\text{P}\left[\exists\mbox{ } a\in\mathbb{R}: \left\Vert M_{n}-a\text{\ensuremath{\mathbb{I}}}_{n}\right\Vert _{F}\le\epsilon\right],
\end{equation}
where $\left\Vert \centerdot\right\Vert _{F}$ is the Frobenius norm. 

The known measure of G(O/U/S)E matrices for general $\beta$ is
\[
\mu_{n}(\beta)=C_{n}(\beta)\text{ }e^{-\frac{\beta}{4}\text{tr}(M_{n})^{2}}dM_{n}.
\]
Comment: The ensemble is clearly invariant under any orthogonal transformation
$M_{n}\rightarrow UM_{n}U^{-1}$, where $U$ is a unitary ($\beta-$orthogonal
in general).

Take the eigenvalue decomposition $M_{n}=UDU^{-1}$ with $D=\text{diag}(\lambda_{0},\lambda_{1},\dots,\lambda_{n-1})$
and $\lambda_{0}\le\lambda_{1}\le\dots\le\lambda_{n-1}$ being the
eigenvalues. 

Fix $a\in\mathbb{R}$, and denote by $\mathbb{I}_{n}$ the $n\times n$
identity matrix. Let $0<\epsilon\ll1$ and let us compute the probability
\[
\text{P}\left[\left\Vert M_{n}-a\mathbb{I}_{n}\right\Vert _{F}\le\epsilon\right].
\]
The probability density of $M_{n}$ is proportional to
\[
e^{-\frac{\beta}{4}\text{Tr}(a\mathbb{I}_{n})}=e^{-\frac{\beta na^{2}}{4}}
\]
and the volume of a ball of radius $\epsilon$ in $n^{2}-$dimensions
is proportional to $\epsilon^{n^{2}}$, so \cite[Section 2.6]{tao2012topics}
\begin{equation}
\text{P}\left[\left\Vert M_{n}-a\mathbb{I}_{n}\right\Vert _{F}\le\epsilon\right]=\left(C'_{n}(\beta)+o(1)\right)\epsilon^{n^{2}}e^{-\frac{\beta na^{2}}{4}},\label{eq:Prob_near_a-1}
\end{equation}
where $C'_{n}(\beta)$ is a constant that depends on $n$ and $\beta$
and $o(1)$ term goes to zero as $\epsilon$ goes to zero. 

This serves as a lower bound for Eq. \eqref{eq:ProblemStatement}.
But $a$ is an arbitrary point on the real line, the contribution
from all such points is upper-bound by the integral
\begin{equation}
\int_{-\infty}^{\infty}\text{P}\left[\left\Vert M_{n}-a\mathbb{I}_{n}\right\Vert _{F}\le\epsilon\right]da=\left(C''_{n}(\beta)+o(1)\right)\sqrt{\frac{\pi}{n\beta}}\text{ }\epsilon^{n^{2}},\label{eq:Final-1}
\end{equation}
for a new constant $C''_{n}(\beta)$. Since $n$ is fixed, we conclude
that
\begin{equation}
\text{P}\left[\exists\mbox{ } a\in\mathbb{R}: \left\Vert M_{n}-a\text{\ensuremath{\mathbb{I}}}_{n}\right\Vert _{F}\le\epsilon\right]=\left(C'''_{n}(\beta)+o(1)\right)\epsilon^{n^{2}},\label{eq:FinalResult-1}
\end{equation}
where $C'''_{n}(\beta)$ is yet a new constant. 

Suppose the number of overlapping terms with $H_{p,q}$ is $z$. Because
of the independence, the probability that all of them are $\epsilon-$close
to a constant multiple of the identity is (we restore $n=d^{2}$)
\[
\left(K_{n}(\beta)+o(1)\right)\epsilon^{zd^{4}}
\]
where $K_{n}(\beta)$ is a constant. Moreover the probability that
the smallest two eigenvalues of a matrix chosen from the Gaussian
ensemble are $\epsilon$-close is bounded by $\epsilon^{4}$ \cite{erdHos2009local}.
Therefore, the probability that a rare local region given by Eq. \eqref{eq:rareLocal1}
occurs is
\[
\left(K_{d}'(\beta)+o(1)\right)\epsilon^{zd^{4}+4}.
\]
Note that the $\beta$ dependence only enters through the pre-factor
which is independent of $\epsilon$. 

How large does the system have to be for the expected gap to be as
small as $\epsilon?$ For the gap to be $\epsilon$ small, the expected
number of term in the Hamiltonian Eq. \eqref{eq:H} is
\[
N\sim\epsilon^{-zd^{4}-4}.
\]
\subsection{Proof of Corollary \ref{cor:discreteEigs}}
\begin{proof}
If the eigenvalue distribution is discrete, in the proof of Theorem
\ref{thm:Id_eigenvals} above, we would have a finite probability
of existence of a $H_{p,q}$ whose smallest eigenvalue is \textit{exactly}
$k-$fold degenerate and its nearest neighbors are exactly proportional
to the identity matrix. Let the ground state of all the terms in the
Hamiltonian excluding $H_{p,q}$ be $|\theta_{E}\rangle$, and let
us denote the eigenvectors of the smallest eigenvalue of $H_{p,q}$
by $|\psi_{p,q}^{(\ell)}\rangle$ where $\ell\in\{1,\dots,k\}$. Then
the ground states of $H$ are product states $|\theta_{E}\rangle\otimes|\psi_{p,q}^{(\ell)}\rangle$
and are $k-$fold degenerate. 
\end{proof}
\subsection{Proof of Theorem 2}
\begin{proof}
Let an arbitrary vertex on the lattice or the graph be at the fixed
position $\mathbf{r_{0}}$ and rewrite Eq. \eqref{eq:H} as 
\begin{equation}
H=\sum_{|\langle i,\mathbf{r_{0}}\rangle|=1}\mathbb{I}\otimes H_{i,\mathbf{r_{0}}}+\sum_{|\langle i,j\rangle|\ge2}\mathbb{I}\otimes H_{i,j}\label{eq:HammingDistance}
\end{equation}
where the first sum includes all the terms that act on $\mathbf{r_{0}}$
and its neighboring vertices (distant $1$), and the second sum includes
the rest of the interactions.

To prove the first statement take $H_{i,j}$ in Eq. \eqref{eq:H}
to be random projectors with the rank $d(d-1)$. Let $\pi=\mbox{diag}(0,1,\dots,1)$
be the diagonal projector of rank $d-1$ and $\mathbb{I}_{\mathbf{r_{0}}}$
an $d\times d$ identity matrix acting on $\mathbf{r_{0}}$. Since
eigenvectors are Haar, there is a positive probability density that
all the terms in the first sum in Eq. \eqref{eq:HammingDistance}
take the form $\pi\otimes\mathbb{I}_{\mathbf{r_{0}}}$. Examples of
this are illustrated in Fig. \eqref{fig:An-example:-Configuration}.

By the independence of local terms and Haar-distribution of the local eigenvectors,
for sufficiently large $N$ almost surely there will be a site $\mathbf{r_{0}}$,
whose neighbors are $\epsilon$ close to projectors of the form just
described. By assumption $\epsilon=0$ corresponds to neighbors having
the exact form.

We write 
\begin{align}
\sum_{|\langle i,\mathbf{r_{0}}\rangle|=1}\mathbb{I}\otimes H_{i,\mathbf{r_{0}}} & =\sum_{|\langle i,\mathbf{r_{0}}\rangle|=1}\mathbb{I}\otimes(H_{i,\mathbf{r_{0}}}^{0}+\delta H_{i,\mathbf{r_{0}}}),\label{eq:Decompose_comm_}
\end{align}
where the superscript zero denotes $\epsilon=0$, and by assumption
$H_{i,\mathbf{r_{0}}}^{0}=\pi_{i}\otimes\mathbb{I}_{\mathbf{r_{0}}}$
and $||\mathbb{I}\otimes\delta H_{i,\mathbf{r_{0}}}||\le\epsilon$.

We define the $\epsilon=0$ Hamiltonian by $H_{0}\equiv\sum_{|\langle i,\mathbf{r_{0}}\rangle|=1}\mathbb{I}\otimes H_{i,\mathbf{r_{0}}}^{0}+\sum_{|\langle i,j\rangle|\ge2}\mathbb{I}\otimes H_{i,j}$,
which is equal to 
\begin{eqnarray}
H_{0} & = & \mathbb{I}_{\mathbf{r_{0}}}\otimes\{\sum_{|\langle i,\mathbf{r_{0}}\rangle|=1}\pi_{i}\otimes\mathbb{I}+\sum_{|\langle i,j\rangle|\ge2}\mathbb{I}\otimes H_{i,j}\}\nonumber \\
 & \equiv & \mathbb{I}_{\mathbf{r_{0}}}\otimes H_{E},\label{eq:H0}
\end{eqnarray}
where $H_{E}$ is the Hamiltonian (inside braces) acting on all sites other than $\mathbf{r_{0}}$
and $\mathbb{I}$ in the distant-2 terms excludes
$i,j$ and $\mathbf{r_{0}}$. By eigenvalue decomposition we have
$H_{E}=\Theta_{E}^{-1}\Lambda_{E}\Theta_{E}$, where $\Lambda_{E}$
is the diagonal matrix of eigenvalues and $\Theta_{E}$ is the unitary
matrix of eigenvectors. We arrive at the desired result where $H_{0}=\mathbb{I}_{r_{0}}\otimes\Theta_{E}^{-1}\Lambda_{E}\Theta_{E}$;
the eigenvalues of this Hamiltonian all have a $d-$fold algebraic
multiplicity.

Let $|\theta_{E}\rangle$ denote the ground state of $H_{E}$ with
the energy $\lambda_{E}\ge0$, the ground states of $H_{0}$ can be
taken to be 
\begin{equation}
|\Psi_{k}\rangle=|e_{k}\rangle\otimes|\theta_{E}\rangle,\label{eq:psi_unperturbed}
\end{equation}
where $|e_{k}\rangle$ are the standard vectors supported at $\mathbf{r_{0}}$
and the corresponding eigenvalues of $H_{0}$ are $\lambda_{min}^{0,(k)}=\lambda_{E}$,
where $k\in\{1,2,...,d\}$. For $\epsilon>0$ what used to be degenerate
ground state energies of $H_{0}$, generically split, i.e., $\lambda_{min}^{\epsilon,(k)}\ne\lambda_{min}^{\epsilon,(\ell)}$,
where $k,\ell\in\{1,\dots,d\}$ and $k\ne\ell$. Since $||\sum_{|\langle i,\mathbf{r_{0}}\rangle|=1}\mathbb{I}\otimes\delta H_{i,\mathbf{r_{0}}}||\le z||\delta H_{max}||$,
Weyl inequalities imply that for
all $k$ 
\begin{eqnarray}
-z\epsilon\le\lambda_{min}^{\epsilon,(k)}-\lambda_{E}^{(0)} & \le & z\epsilon\label{eq:projectors}
\end{eqnarray}
where, as before, $||\mathbb{I}\otimes\delta H_{i,\mathbf{r_{0}}}||=||\delta H_{i,\mathbf{r_{0}}}||$,
$z$ is the number of nearest neighbor terms to $\mathbf{r_{0}}$
(e.g., $z=2D$ on a square lattice in $D$ spatial dimensions) and
$\delta H_{max}$ is the overlapping term with the maximum error norm.

Since the eigenvectors are Haar and the dimension of the local kernels
is $d$, for any fixed $\epsilon$ we can find a sufficiently large
$N$ in which there is a site whose all neighbors are $\epsilon$
close to $\mathbb{I}_{\mathbf{r_{0}}}\otimes\pi$. Since $z$ is finite
and $\epsilon$ can is arbitrary, by Eq. \eqref{eq:projectors}
the $d$ smallest eigenvalues of $H_{0}$ will be $\mathcal{O}(\epsilon)$
apart resulting in an arbitrary small gap.

The argument formally generalizes to local projectors with
ranks $d(d-h)$, with $d>h\ge1$. All that needs to be changed in
the proof is to let $2\le\mbox{dim}(\mbox{Ker}(\pi))<d$. 

Lastly if we allow variable ranks of $H_{i,j}$, then the same construction
as above guarantees having a site on which the Hamiltonian acts trivially
and the above argument guarantees $\epsilon$ splitting of the eigenvalues
and lack of an energy gap in the limit. 
\end{proof}
\subsection{Proof of Corollary \ref{cor:GAPdiscreteEigs-1}}
\begin{proof}
Suppose $\lambda_{pq}^{1},\lambda_{pq}^{2},\dots,\lambda_{pq}^{d}$
are $d$ discrete eigenvalues of the distribution that are not all
equal. Then with an arbitrarily high probability we can find a site,
$\mathbf{r_{o}}$, whose neighbors are all $\epsilon$ close to having
the form $\mathbb{I}_{\mathbf{r_{o}}}\otimes\mbox{diag}(\lambda_{pq}^{1},\lambda_{pq}^{2},\dots,\lambda_{pq}^{d})$.
In Theorem \ref{thm:Projectors} replacing $\pi$ with $\mbox{diag}(\lambda_{pq}^{1},\lambda_{pq}^{2},\dots,\lambda_{pq}^{d})$
and following the same argument that lead to Eq. \eqref{eq:projectors},
we arrive at the bound on the gap being $|\lambda_{min}^{\epsilon,(k)}-\lambda_{E}^{(0)}|\le2z\epsilon$,
where $\epsilon$ can be arbitrary small. 
\end{proof}

\end{document}